\documentclass[1opt,twocolumn,letter]{IEEEtran}
\usepackage[dvips]{graphicx}
\DeclareGraphicsExtensions{.eps}
\graphicspath{{figures/}}
\usepackage[tight]{subfigure}
\usepackage{fancyhdr}
\usepackage{hyperref}
\usepackage{wrapfig}
\usepackage{amsmath}
\usepackage{mathrsfs}
\usepackage{amsfonts}
\input{notations.sty}
\usepackage[ruled,slide, vlined]{algorithm2e}
\usepackage{multicol}
\usepackage{caption}

\usepackage{tikz}
\usetikzlibrary{arrows,automata}
\usetikzlibrary{shapes,snakes,backgrounds,petri,calc}
\usetikzlibrary{positioning}

\usepackage{rotating}

\usepackage{times}

\newcommand{\ie}{\textit{i.e.},\xspace}

\newcommand{\ignore}[1]{{}}

\makeatletter \newcommand \listoftodos{\section*{Todo list} \@starttoc{tdo}}
\newcommand\l@todo[2]{
  \par\noindent \textit{#2}, \parbox{10cm}{#1}\par
} 
\newcommand{\ordi}[3]{\draw[#3] ( #1, #2 ) -- ( #1+1, #2 ) -- ( #1+1, #2+1 )  -- ( #1, #2+1 ) -- ( #1, #2) -- ( #1-.3, #2-.3) -- ( #1+.7, #2-.3 ) -- ( #1+1, #2 ) ;}

\newcommand{\bonhomme}[3] {%
\draw [#3] ( #1, #2 ) circle ( .3 );%
\draw [#3] ( #1, #2-.3 ) -- ( #1, #2-1 );%
\draw [#3] ( #1-.3, #2-.6 ) -- ( #1+.3, #2-.6);%
\draw [#3] ( #1, #2-1 ) -- ( #1-.3, #2-1.6 );%
\draw [#3] ( #1, #2-1 ) -- ( #1+.3, #2-1.6 );%
}

\newcommand{\server}[3]{%
\draw[#3] ( #1, #2) -- ( #1, #2-2 ) -- ( #1+1, #2-2 ) -- ( #1+1, #2 ) -- ( #1, #2 ) -- ( #1+.3, #2+.3 ) -- ( #1+1.3, #2+.3 ) -- ( #1+1.3, #2-1.7) -- ( #1+1, #2-2 );%
\draw[#3] ( #1+1, #2 ) -- ( #1+1.3, #2+.3 ) ;%
}

\makeatother

\begin{document}

\title{Time Petri Net Models for a New Queuless and Uncentralized
  Resource Discovery System}

\author{
\begin{tabular}{ccc}
Camille Coti & Sami Evangelista & Kais Klai\\
\end{tabular}\\
\textit{Universit\'e Paris 13, Sorbonne Paris Cit\'e, LIPN, CNRS, UMR 7030\\
  F-93430, Villetaneuse, France}\\
\textit{\{first.last\}@univ-paris13.fr}
}

\date{} 

\maketitle
\begin{abstract}
In this report, we detail the model using Petri Nets of a new fully
distributed resource reservation system. The basic idea of the
considered distributed system is to let a user reserve a set of
resources on a local network and to use them, without any specific,
central administration component such as a front-end node. Resources
can be, for instance, computing resources (cores, nodes, GPUs...) or
some memory on a server. In order to verify some qualitative and
quantitative properties provided by this system, we need to model
it. We detail the algorithms used by this system and the Petri Net
models we made of it. 
\end{abstract}

\section{A New Fully Distributed Ressource Management System}
\label{sec:algo}

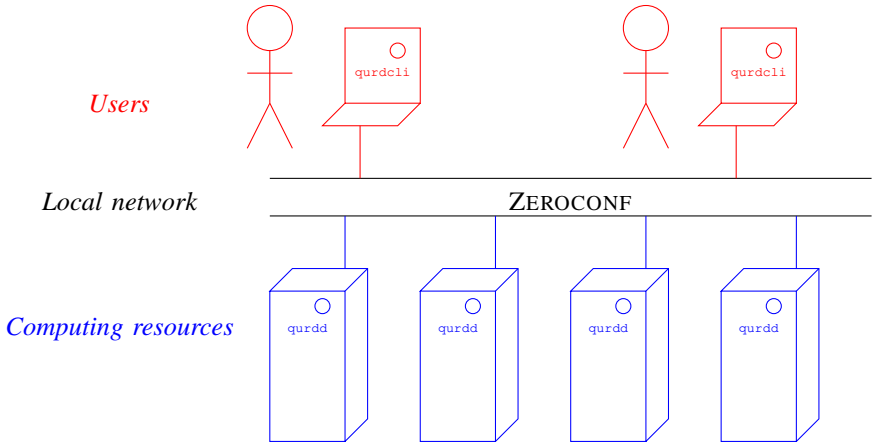
\begin{figure*}
\begin{tikzpicture}


\draw ( 0, 0 ) -- ( 8, 0 );
\draw ( 0, -.5 ) -- ( 8, -.5 );
\node at ( 4, -.3) {Z{\sc eroconf}};
\node at ( -2, -.3) {\it Local network};


\ordi{1}{1}{color=red}
\draw [color=red] ( 1.2, .7 ) -- ( 1.2, 0 );
\bonhomme{0}{2}{color=red}
\draw [color=red] ( 1.7, 1.7 ) circle ( .1 );
\node[color=red] at ( 1.5, 1.4) {\tiny \tt qurdcli};

\ordi{6}{1}{color=red}
\draw [color=red] ( 6.2, .7 ) -- ( 6.2, 0 );
\bonhomme{5}{2}{color=red}
\draw [color=red] ( 6.7, 1.7 ) circle ( .1 );
\node[color=red] at ( 6.5, 1.4) {\tiny \tt qurdcli};

\node [color=red] at ( -2, 1) {\it Users};


\server{0}{-1.5}{color=blue}
\draw [color=blue] ( 1, -1.2 ) -- ( 1, -.5 );
\draw [color=blue] ( .7, -1.7 ) circle ( .1 );
\node[color=blue] at ( .5, -2) {\tiny \tt qurdd};

\server{2}{-1.5}{color=blue}
\draw [color=blue] ( 3, -1.2 ) -- ( 3, -.5 );
\draw [color=blue] ( 2.7, -1.7 ) circle ( .1 );
\node[color=blue] at ( 2.5, -2) {\tiny \tt qurdd};

\server{4}{-1.5}{color=blue}
\draw [color=blue] ( 5, -1.2 ) -- ( 5, -.5 );
\draw [color=blue] ( 4.7, -1.7 ) circle ( .1 );
\node[color=blue] at ( 4.5, -2) {\tiny \tt qurdd};

\server{6}{-1.5}{color=blue}
\draw [color=blue] ( 7, -1.2 ) -- ( 7, -.5 );
\draw [color=blue] ( 6.7, -1.7 ) circle ( .1 );
\node[color=blue] at ( 6.5, -2) {\tiny \tt qurdd};

\node [color=blue] at ( -2, -2) {\it Computing resources};

\end{tikzpicture}
\caption{\label{fig:archi}Architecture of a QURD system.}
\end{figure*}

In this section, we describe how the machines are reserved for a
user. Our system is made of two parts: the \emph{launcher} (called
{\it qurdcli}, as ``qurd client''), which is executed by the user who
wants to run a job on a set of computing nodes, and the \emph{agent}
(called {\it qurdd}, as ``qurd daemon''), which is a daemon running on
all the resources that exist in the system. This architecture is
depicted in figure \ref{fig:archi}.

Our algorithm relies on the service discovery tools provided by the
Zeroconf protocol \cite{zeroconf}. 
Computing nodes declare themselves on the Zeroconf bus,
like any network service. Then a user's launcher can look on the local
Zeroconf bus which nodes are available. However, this simple discovery
service is not sufficient to ensure that the computing resources will
not be used by several jobs at the same time.
\subsection{The Job Launcher}
\label{sec:algo:launcher}

The launcher's behavior is quite straightforward (see Algorithm~\ref{algo:globalsub}): it reserves nodes
and, as soon as it has enough nodes for its job, it starts the job on
the said nodes. 

\begin{algorithm}[ht]
  \KwData{Job J to execute on N nodes}
  M = reserve\_nodes( N )\;
  \If{ card( M ) == N } {
    start\_job( J, M )\;
  }
  \caption{\label{algo:globalsub} Job start-up general algorithm}
\end{algorithm}


The reservation algorithm uses the service discovery features of
Zeroconf. It listens on the Zeroconf bus and finds out which computing
nodes have published themselves (as available). The information published by each
node includes the host's name and a port on which the node can be
contacted. Then, for each node it has discovered, the launcher contacts
the machine on the declared port. If the node is available, the
launcher receives an acknowledgement ("OK''); otherwise it receives a
rejection ("KO''). In the former case, it inserts the node on its set
of reserved nodes. As soon as the set of reserved nodes contains the
required number of computing nodes, it starts launching the job on
these nodes. In case there is not enough available resources, we can define two different
behaviors: either the reservation returns the (smaller) set of
resources it was able to reserve, or it waits until it gets all the
necessary resources (see Algorithm~\ref{algo:reservationWait}).

\begin{algorithm}[ht]
  \SetFuncSty{textbf}
  \SetKwFunction{reservenodes}{reserveNodes}
  \reservenodes{ nbNodes } 
  \Begin{
      \KwData{machines = \{\}}
      listenZeroconf()\;
      \ForEach{machine $m$ newly discovered} {
        \If{ card( machines ) $<$ nbNodes } {
          contactMachine( $m$ ) \;
          ack = receiveAck( $m$ )\;
          \If{ ack == OK } {
            machines.append( $m$ ) \;
          }
        }
      }
      \eIf{ card( machines ) == nbNodes } {
        \Return{machines} \;
      } {
        freeMachines( machines ) \;
        \Return{\{\}} \;
      }
      
    }
\caption{\label{algo:reservationFail}Resource reservation algorithm (fail semantics)}
\end{algorithm}

\begin{algorithm}[ht]
  \SetFuncSty{textbf}
  \SetKwFunction{reservenodes}{reserveNodes}
  \reservenodes{ nbNodes } 
  \Begin{
      \KwData{machines = \{\}}
      \While{ card( machines ) $<$ nbNodes } {
        listenZeroconf()\;
        \ForEach{machine $m$ newly discovered} {
          \If{ card( machines ) $<$ nbNodes } {
            contactMachine( $m$ ) \;
            ack = receiveAck( $m$ )\;
            \If{ ack == OK } {
              machines.append( $m$ ) \;
            }
          }
        }
      }
      \Return{machines} \;
    }
\caption{\label{algo:reservationWait}Resource reservation algorithm (wait semantics)}
\end{algorithm}

\subsection{The Computing Resources (Machines)}
\label{sec:algo:machine}

When it is available (\ie it can be used to run a job), a machine
publishes itself on the Zeroconf bus. When it switches to the unavailable state,
it unpublishes itself. However, due to the latency of Zeroconf and to a possible concurrency between
the users, the machines also need to implement a local reservation system to
make sure that they are used by only one job at a given moment. 

Each machine executes Algorithm~\ref{algo:resource} in order to
interact with clients and to execute their jobs when possible. A
machine can be in one of the three following states: \emph{available}
(the machine is free, it can be reserved for a job); \emph{reserved}
(the machine is reserved for a job but the job is not running on it
yet); \emph{running} (a job is running on the node). 

%

%

When a machine is in the state \emph{available}, it can be reserved
by a user. The user is then considered by the machine as its
\emph{client}. Upon request reception, it sends an acknowledgement to
the client, switches into \emph{reserved} state and unpublishes itself
from Zeroconf. If a machine receives a request from a client while it
is not in \emph{available} state, it sends a rejection to the client.  When a machine is in the state \emph{reserved}, it can receive a job from its client. The machine executes this job until completion. Then it
switches into the state \emph{available} and publishes itself back on Zeroconf. 

It can happen that the client does not obtain the required number of machines. In this case, when at state \emph{reserved}, a machine can be released which makes it switch to the state \emph{available} (the machine publishes itself back on Zeroconf).

\begin{algorithm}[ht]
  \SetFuncSty{textbf}
  \SetKwFunction{waitForAJob}{waitjob}
  \waitForAJob{ }
  \Begin{
      \KwData{state = available}
      \KwData{client = -1}
      publishMyselfOnZeroconf() \;
      \While{True} {
        received, c = recvFromClient() \;
        \Switch{ received.type }{
          \Case{ reservation } {
            \eIf{state == reserved } {
              c.send( KO ) \;
            }{
              c.send( OK ) \;
              state = reserved \;
              client = c \;
              unpublishFromZeroconf() \;
            }
          }
          \Case{job } {
            \eIf{state == reserved OR c != client } {
              refuse() \;
           } {
             executeJob( received ) \;
             state = running \;
             publishMyselfOnZeroconf() \;
             state = available \;
             client = -1 \;
            }
          }
          \Case{ release } {
            publishMyselfOnZeroconf() \;
            state = available \;
            client = -1 \;
          }
        }
      }
    }
    \caption{\label{algo:resource}Algorithm executed on each resource}
\end{algorithm}

\section{Petri Nets-based Model of The Ressources Reservation System}
\label{sec:model}


In the following, we consider that an \emph{application} (or a
\emph{job}) is made of several \emph{processes} that are meant to run
on a set of \emph{resources}, also called \emph{machines}. The user
submits an application through a \emph{client}. For the sake of readability, we describe the model of each component of the system separately. The reader can find in the annex a full model involving one application and four ressources. A couloured Petri net model is presented as well allowing to have more compact and parameterized model.

\begin{sidewaysfigure*}[ht!]
  \begin{center}
    \resizebox{0.8\textwidth}{!}{
\usetikzlibrary{arrows,automata}
\usetikzlibrary{shapes,snakes,backgrounds,petri,calc}
\usetikzlibrary{positioning}

\begin{tikzpicture}[node distance=1.3cm,>=stealth',auto]

  \node [place, tokens=1, label=begin] (begin0) {};
  \node [transition] (start0) [below = of begin0, label=right:start\_job] {};
  \node [place, below =of start0, label=right:get\_nodes] (getnodes0) {};

  \draw [->] (begin0) -- (start0);
  \draw [->] (start0) -- (getnodes0) node [midway] {$n$};

  \node [place, below =of getnodes0, label=above:answered] (answered0) {};
  \node [transition] (launch0) [below= of answered0, label=above:launch] {};
  \node [place, below =of launch0, label= left:launching\_job] (launching0) {};
  \draw [->] (answered0) -- (launch0) node [midway] {$n$};
  \draw [->] (launch0) -- (launching0) node [midway] {$n$};


  \node [place, tokens=1, label=available, below left= of launching0, xshift = -2.5cm] (available1) {};
  \node [transition] (t11) [below = of available1, label=right:t1] {};
  \node [place, tokens=0, label=below right:reserved, below of=t11] (reserved1) {};
  \node [transition] (t21) [below of=reserved1, label=right:t2] {};
  \node [place, tokens=0, label=right:running, below of=t21] (running1) {};
  \node [transition] (t31) [below of=running1,label=right:t3] {};
  \node [place, tokens=0, label=right:finished, below of=t31] (finished1) {};
  \node [transition] (t41) [left=of finished1, label=left:t4] {};
  \node [transition] (t51) [right=of reserved1, label=right:cancel] {};

  \draw [->] (available1) -- (t11);
  \draw [->] (t11) -- (reserved1);
  \draw [->] (reserved1) -- (t21);
  \draw [->] (t21) -- (running1);
  \draw [->] (running1) -- (t31);
  \draw [->] (t31) -- (finished1);
  \draw [->] (finished1) -- (t41);
  \draw [->] (t41) -- (available1);
  \draw [->] (reserved1) -- (t51);
  \draw [->] (t51) -- (available1);

    \draw [->] (getnodes0) -- (t11);
  \draw [->] (launching0) -- (t21);
  \draw [->] (t11) -- (answered0);
  \draw [->] (t51) -- (getnodes0);
  \draw [->] (answered0) -- (t51);

\node [right=of t51.west |- available1.north] (T1) {};
\node [left=of finished1.south -| t41.east] (B1) {};
    \draw[dashed, draw] ($(T1) + (.5, .5)$) rectangle ($(B1)+(.3, -.2)$); 


  \node [place, tokens=1, label=available, below right = of launching0, xshift = 2.5cm] (available2) {};
  \node [transition] (t12) [below = of available2, label=right:t1] {};
  \node [place, tokens=0, label=below right:reserved, below of=t12] (reserved2) {};
  \node [transition] (t22) [below of=reserved2, label=right:t2] {};
  \node [place, tokens=0, label=right:running, below of=t22] (running2) {};
  \node [transition] (t32) [below of=running2,label=right:t3] {};
  \node [place, tokens=0, label=right:finished, below of=t32] (finished2) {};
  \node [transition] (t42) [left=of finished2, label=left:t4] {};
  \node [transition] (t52) [right=of reserved2, label=right:cancel] {};

  \draw [->] (available2) -- (t12);
  \draw [->] (t12) -- (reserved2);
  \draw [->] (reserved2) -- (t22);
  \draw [->] (t22) -- (running2);
  \draw [->] (running2) -- (t32);
  \draw [->] (t32) -- (finished2);
  \draw [->] (finished2) -- (t42);
  \draw [->] (t42) -- (available2);
  \draw [->] (reserved2) -- (t52);
  \draw [->] (t52) -- (available2);

    \draw [->] (getnodes0) -- (t12);
  \draw [->] (launching0) -- (t22);
  \draw [->] (t12) -- (answered0);
  \draw [->] (t52) -- (getnodes0);
  \draw [->] (answered0) -- (t52);

\node [right=of t52.west |- available2.north] (T2) {};
\node [left=of finished2.south -| t42.east] (B2) {};
    \draw[dashed, draw] ($(T2) + (.5, .5)$) rectangle ($(B2)+(.3, -.2)$); 


  \node [place, tokens=1, label=available, left = of available1, xshift=-5cm] (available0) {};
  \node [transition] (t10) [below = of available0, label=right:t1] {};
  \node [place, tokens=0, label=below right:reserved, below of=t10] (reserved0) {};
  \node [transition] (t20) [below of=reserved0, label=right:t2] {};
  \node [place, tokens=0, label=right:running, below of=t20] (running0) {};
  \node [transition] (t30) [below of=running0,label=right:t3] {};
  \node [place, tokens=0, label=right:finished, below of=t30] (finished0) {};
  \node [transition] (t40) [left=of finished0, label=left:t4] {};
  \node [transition] (t50) [right=of reserved0, label=right:cancel] {};

  \draw [->] (available0) -- (t10);
  \draw [->] (t10) -- (reserved0);
  \draw [->] (reserved0) -- (t20);
  \draw [->] (t20) -- (running0);
  \draw [->] (running0) -- (t30);
  \draw [->] (t30) -- (finished0);
  \draw [->] (finished0) -- (t40);
  \draw [->] (t40) -- (available0);
  \draw [->] (reserved0) -- (t50);
  \draw [->] (t50) -- (available0);
  
  \draw [->] (getnodes0) -- (t10);
  \draw [->] (launching0) -- (t20);
  \draw [->] (t10) -- (answered0);
  \draw [->] (t50) -- (getnodes0);
  \draw [->] (answered0) -- (t50);

\node [right=of t50.west |- available0.north] (T0) {};
\node [left=of finished0.south -| t40.east] (B0) {};
    \draw[dashed, draw] ($(T0) + (.5, .5)$) rectangle ($(B0)+(.3, -.2)$); 


  \node [place, tokens=1, label=available, right = of available2, xshift = 5cm] (available3) {};
  \node [transition] (t13) [below = of available3, label=right:t1] {};
  \node [place, tokens=0, label=below right:reserved, below of=t13] (reserved3) {};
  \node [transition] (t23) [below of=reserved3, label=right:t2] {};
  \node [place, tokens=0, label=right:running, below of=t23] (running3) {};
  \node [transition] (t33) [below of=running3,label=right:t3] {};
  \node [place, tokens=0, label=right:finished, below of=t33] (finished3) {};
  \node [transition] (t43) [left=of finished3, label=left:t4] {};
  \node [transition] (t53) [right=of reserved3, label=right:cancel] {};

  \draw [->] (available3) -- (t13);
  \draw [->] (t13) -- (reserved3);
  \draw [->] (reserved3) -- (t23);
  \draw [->] (t23) -- (running3);
  \draw [->] (running3) -- (t33);
  \draw [->] (t33) -- (finished3);
  \draw [->] (finished3) -- (t43);
  \draw [->] (t43) -- (available3);
  \draw [->] (reserved3) -- (t53);
  \draw [->] (t53) -- (available3);
  
  \draw [->] (getnodes0) -- (t13);
  \draw [->] (launching0) -- (t23);
  \draw [->] (t13) -- (answered0);
  \draw [->] (t53) -- (getnodes0);
  \draw [->] (answered0) -- (t53);

\node [right=of t53.west |- available3.north] (T3) {};
\node [left=of finished3.south -| t43.east] (B3) {};
    \draw[dashed, draw] ($(T3) + (.5, .5)$) rectangle ($(B3)+(.3, -.2)$); 


  \node [place, below=of finished1, xshift=2.5cm, label=right:job\_finished] (j_finished0) {};
  \node [transition] (t60) [below = of j_finished0, label=right:t5] {};
  \node [place, below=of t60, label=right:job\_done] (done0) {};

  \draw [->] (t40) -- (j_finished0);
  \draw [->] (t41) -- (j_finished0);
  \draw [->] (t42) -- (j_finished0);
  \draw [->] (t43) -- (j_finished0);

  \draw [->] (j_finished0) -- (t60) node [midway] {$n$};
  \draw [->] (t60) -- (done0);

\end{tikzpicture}	
}
    \caption{\label{fig:full}Complete model for 4 machines and 1 job}
  \end{center}
\end{sidewaysfigure*}
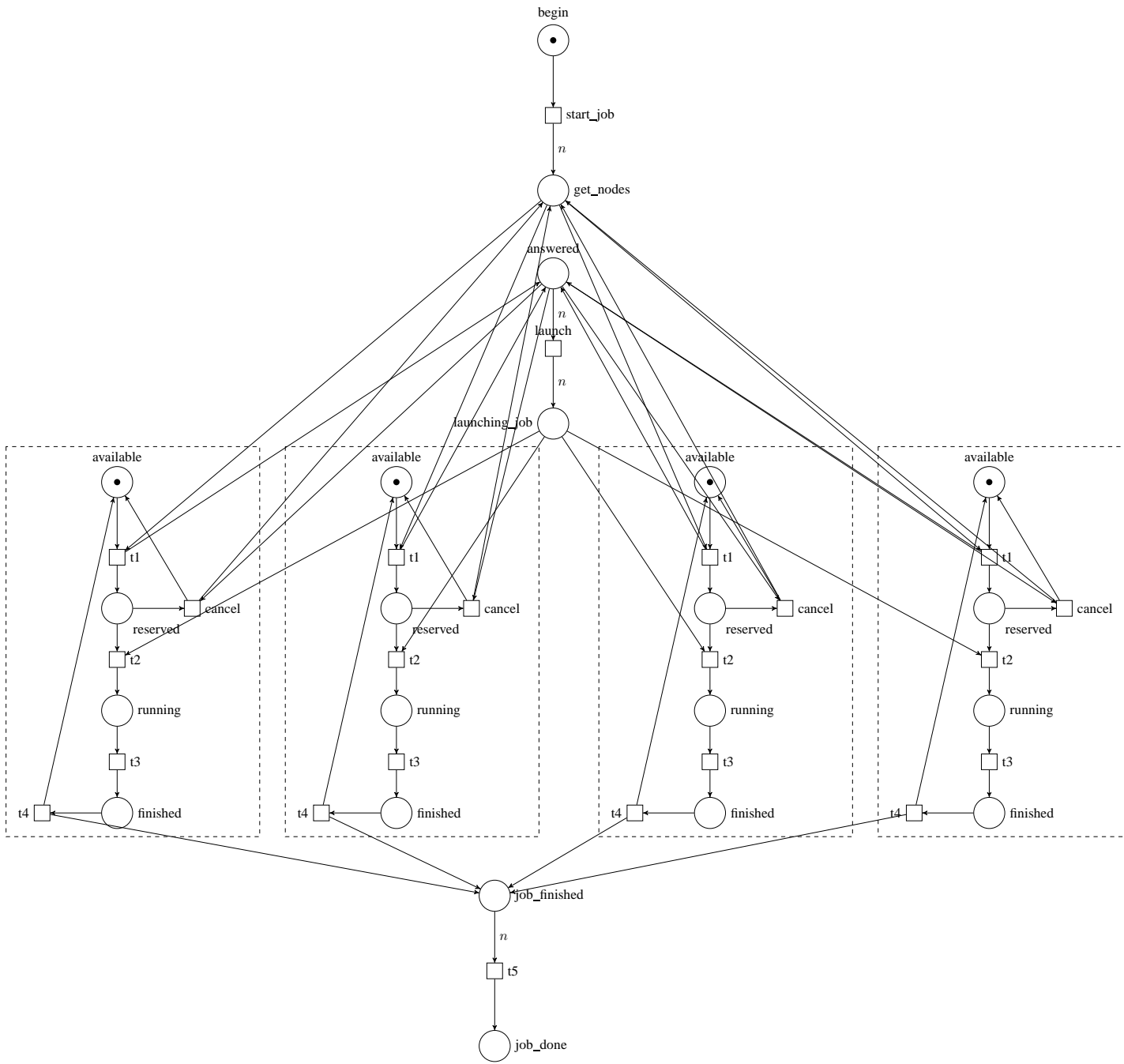

\begin{sidewaysfigure*}[ht!]
  \begin{center}
    \resizebox{0.8\textwidth}{!}{\input{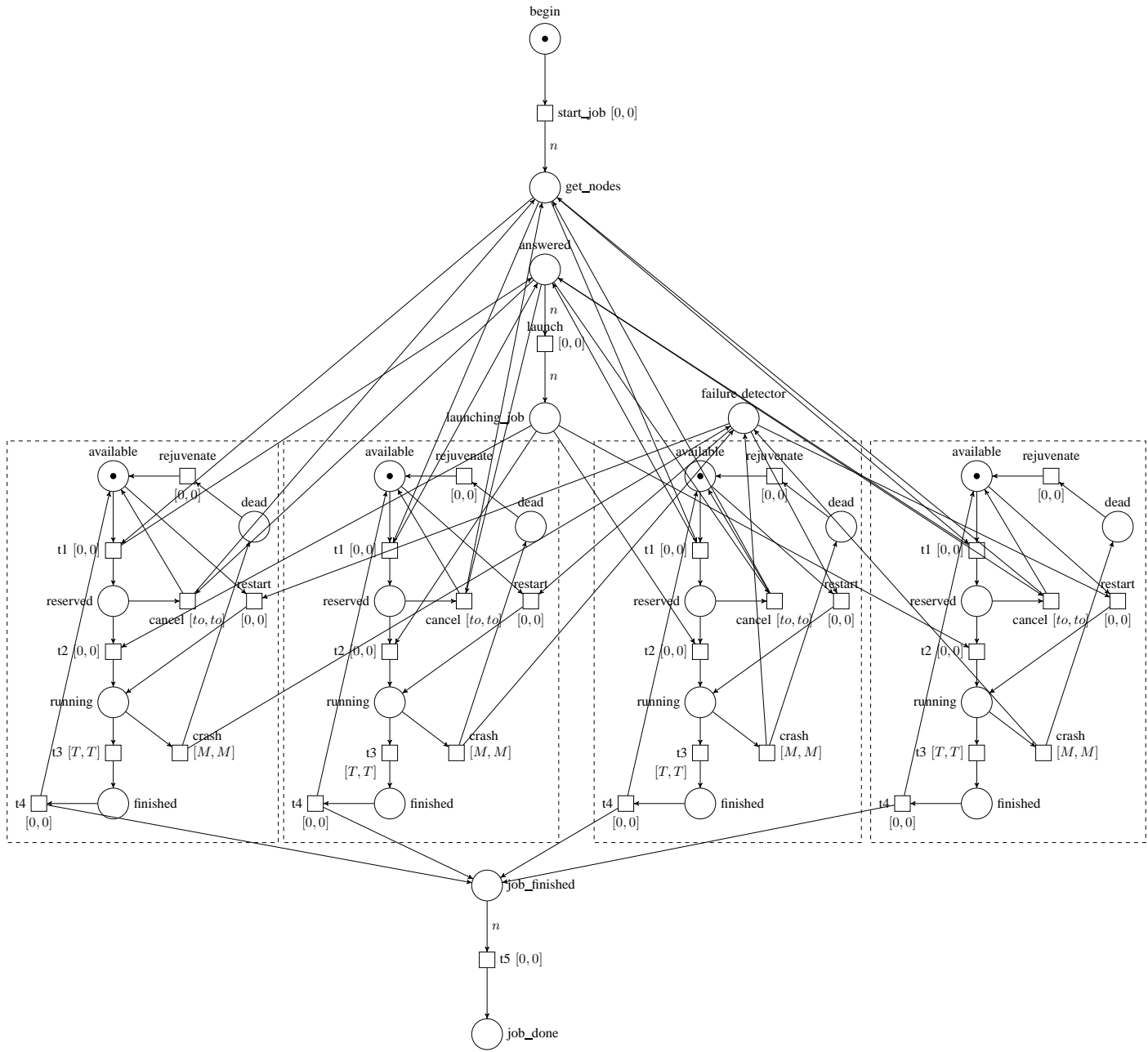}}	
    \caption{\label{fig:fullfd}Complete model for 4 machines and 1 job,
      with a failure detector}
  \end{center}
\end{sidewaysfigure*}

\begin{figure*}[ht!]
  \begin{center}

\scalebox{.8}{\begin{tikzpicture}[node distance=1.8cm]
   \tikzset{
    place/.style={circle,draw,very thick,minimum size=6mm},
    transition/.style={rectangle,draw,very thick,minimum size=4mm,fill=gray!50}
  }
  \tikzstyle{arc}=[>=stealth',thick,->]
  \tikzstyle{marked-place}=[place,fill=blue!60]
  \node [marked-place, label=begin,pin=180:\small$\sum_{j \in jobs} \tuple{j}$] (begin) {};
  \node [transition] (start) [below of=begin, label=right:start\_job] {};
  \node [place, below  of=start, label=right:get\_nodes] (getnodes) {};
  \node [place, label=answered,left of=begin,node distance=5.4cm] (answered) {};
  \node [transition, label=launch,left of=answered,node distance=5.4cm] (launch) {};
  \node [place, label=left:launching\_job,below of=launch] (launchingjob) {};
  \node [transition,left of=getnodes,label=above right:t1,node distance=5.4cm] (t1) [] {};
  \node [marked-place, label=left:available,left of=t1,node distance=5.4cm,pin=225:\small$\sum_{m \in machines} \tuple{m}$] (available) {};
  \node [place, label=below right:reserved, below of=t1] (reserved) {};
  \node [transition,below of=reserved, label=right:t2] (t2) [] {};
  \node [transition,below of=getnodes, label=right:cancel] (cancel) [] {};
  \node [place, label=left:running, below of=t2] (running) {};
  \node [transition,below of=running, label=right:t3] (t3) [] {};
  \node [place, label=right:finished, below of=t3] (finished) {};
  \node [transition,left of=t2, label=left:t4,node distance=5.4cm] (t4) [] {};
  \node [place, label=left:job\_finished,below of=t4] (jobfinished) {};	
  \node [transition, label=left:t5,below of=jobfinished] (t5) {};	
  \node [place, label=left:job\_done,below of=t5] (jobdone) {};
  \node [transition,label=right:crash,right of=running,node distance=2.7cm] (crash) {};
  \node [place,label=right:dead,below of=crash] (dead) {};
  \node [place,label=right:failure\_detector,above of=crash] (failure) {};
  \node [transition,label=left:continue,left of=t2,node distance=2.7cm] (continue) {};

  \draw [arc] (begin) -- node[right] {$\tuple{j}$} (start);
  \draw [arc] (start) -- node[right] {$P'\tuple{j}$} (getnodes);

  \draw [arc] (answered) -- node[above] {$P'\tuple{j}$} (launch);
  \draw [arc] (launch) -- node[left] {$P'\tuple{j}$} (launchingjob);

  \draw [arc] (available) -- node[above] {$\tuple{m}$} (t1);
  \draw [arc] (getnodes) -- node[above] {$\tuple{j}$} (t1);
  \draw [arc] (t1) -- node[right,pos=0.6] {$\tuple{m,j}$} (reserved);
  \draw [arc] (t1) -- node[right] {$\tuple{j}$} (answered);

  \draw [arc] (reserved) -- node[above] {$\tuple{m,j}$} (cancel);
  \draw [arc] (answered) edge[bend left=15] node[above] {$\tuple{j}$} (cancel);
  \draw [arc] (cancel) edge[bend right=8] node[above,pos=0.3] {$\tuple{m}$} (available);
  \draw [arc] (cancel) -- node[right] {$\tuple{j}$} (getnodes);

  \draw [arc] (reserved) -- node[right] {$\tuple{m,j}$} (t2);
  \draw [arc] (launchingjob) -- node[right] {$\tuple{j}$} (t2);
  \draw [arc] (t2) -- node[right] {$\tuple{m,j}$} (running);

  \draw [arc] (running) -- node[right] {$\tuple{m,j}$} (t3);
  \draw [arc] (t3) -- node[right] {$\tuple{m,j}$} (finished);

  \draw [arc] (finished) -- node[left] {$\tuple{m,j}$} (t4);
  \draw [arc] (t4) -- node[left] {$\tuple{m}$} (available);
  \draw [arc] (t4) -- node[left] {$\tuple{j}$} (jobfinished);

  \draw [arc] (jobfinished) -- node[left] {$P'\tuple{j}$} (t5);
  \draw [arc] (t5) -- node[left] {$\tuple{j}$} (jobdone);

  \draw [arc] (running) -- node[above] {$\tuple{m,j}$} (crash);
  \draw [arc] (crash) -- node[left] {$\tuple{m}$} (dead);
  \draw [arc] (crash) -- node[left] {$\tuple{j}$} (failure);

  \draw [arc] (available) -- node[left] {$\tuple{m}$} (continue);
  \draw [arc] (failure) edge[bend left=11] node[below,pos=0.7] {$\tuple{j}$} (continue);
  \draw [arc] (continue) -- node[left] {$\tuple{j}$} (running);
\end{tikzpicture}}
   \caption{\label{fig:colored}Colored model}
  \end{center}
\end{figure*}
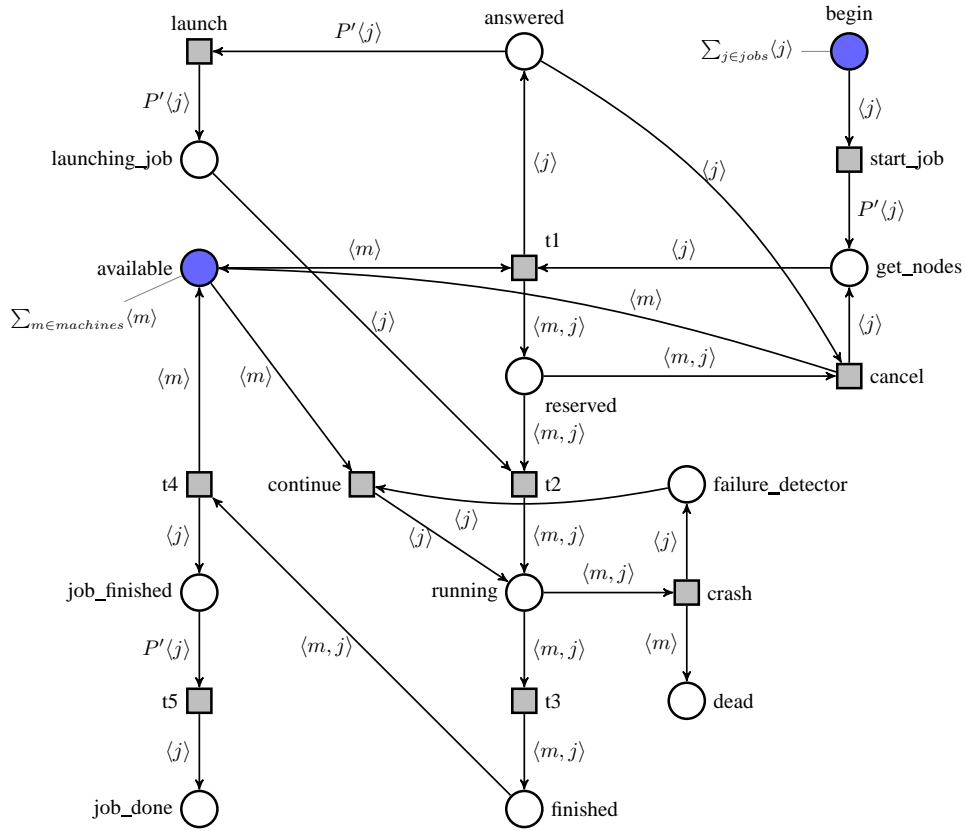

We first describe the model of a computing resource's behaviour in section
\ref{sec:model:qurdd}, and the model of  a
client's behavior in section \ref{sec:model:qurdcli}. We present an optional
model of Zeroconf's semantics in section
\ref{sec:model:zeroconf}. Then we explain how volatility is handled in
section \ref{sec:model:volatile} and how concurrency between jobs on
a given resource is modeled in our system in section
\ref{sec:model:concurrent}. The full models are given in the next
section of this paper, in figure \ref{fig:full}, \ref{fig:fullfd} and
\ref{fig:colored}.

\subsection{Model of a machine: qurdd}
\label{sec:model:qurdd}

\begin{center}
  \scalebox{.6}{\begin{tikzpicture}[node distance=1.3cm,>=stealth',auto]
  \node [place, tokens=1, label=available] (available0) {};
  \node [transition] (t10) [below = of available0, label=right:t1, label=below right:{$[0,0]$}] {};
  \node [place, tokens=0, label=below right:reserved, below of=t10] (reserved0) {};
  \node [transition] (t20) [below of=reserved0, label=right:{t2 $[0,0]$}] {};
  \node [place, tokens=0, label=right:running, below of=t20] (running0) {};
  \node [transition] (t30) [below of=running0,label=right:{t3 $[T,T]$}] {};
  \node [place, tokens=0, label=right:finished, below of=t30] (finished0) {};
  \node [transition] (t40) [left=of finished0, label=left:t4, label=below:{$[0,0]$}] {};
  \node [transition] (t50) [right=of reserved0, label=right:{cancel $[to,to]$}] {};
 
  \draw [->] (available0) -- (t10);
  \draw [->] (t10) -- (reserved0);
  \draw [->] (reserved0) -- (t20);
  \draw [->] (t20) -- (running0);
  \draw [->] (running0) -- (t30);
  \draw [->] (t30) -- (finished0);
  \draw [->] (finished0) -- (t40);
  \draw [->] (t40) -- (available0);
  \draw [->] (reserved0) -- (t50);
  \draw [->] (t50) -- (available0);
  
  \node (res0) [left = of available0]{ \it request};
  \draw [->, dashed] (res0) -- (t10);
  \node (done0) [below = of finished0]{ \it job\_finished};
  \draw [->, dashed] (finished0) -- (done0);
  \node (answ0) [right = of available0]{ \it answered};
  \draw [->, dashed] (t10) -- (answ0);
  \draw [->, dashed] (answ0) -- (t50);
  \node (launch0) [left = of reserved0]{ \it launch\_job};
  \draw [->, dashed] (launch0) -- (t20);

\end{tikzpicture}	}
  
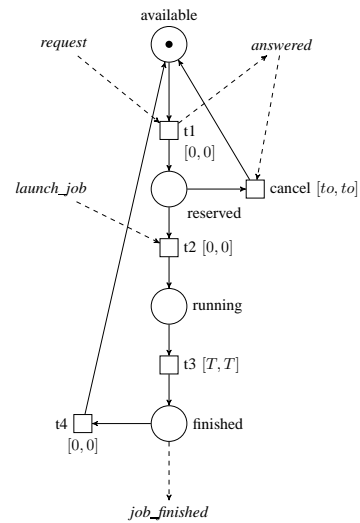
\captionof{figure}{\label{fig:machine} Petri net model of a machine (qurdd daemon)}
\end{center}

The  Petri net model of a machine (figure \ref{fig:machine}) is guided by Algorithm~\ref{algo:resource}. A machine
can be reserved when it is  \emph{available}. It answers the client and switches into
\emph{reserved} mode. Starting from this state, two behaviors are possible: either the reservation is cancelled (through transition \emph{cancel} that can be fired after some timeout) and the machine becomes available again (through a token present at place \emph{available}), or it starts executing the required application and switches to the state \emph{running} (through the firing of transition \emph{t2}). When this local process is done, the machine switches to state \emph{finished}, signals to
the client that its part of the job is done, and then returns back to state \emph{available} (through transition \emph{t4}).

\subsection{Model of the reservation system: qurdcli}
\label{sec:model:qurdcli}

Figure \ref{fig:qurdcli} illustrates the reservation system of
a client (Algorithm~\ref{algo:reservationFail}). For the sake of simplicity and readability, we replaced
the machines by dashed boxes, each corresponds to an
instance of the model represented by figure \ref{fig:machine}. When an application needs to get $n$ resources, the \emph{begin} place triggers a \emph{start\_job} transition whose firing produces $n$ tokens in the
\emph{get\_nodes} place. Each of these tokens corresponds to a
required resource. The reservation system gets a list of the resources
that are declared on the Zeroconf bus. The \emph{get\_nodes} place is
an input place of all the \emph{t1} transitions of the machines that are
present on the Zeroconf bus (see Figure~\ref{fig:machine}). If the
machine is actually available, a token is present in the
place \emph{available} and the transition \emph{t1} can be fired leading to the reservation of the machine by this application (place \emph{reserved} is marked). Also, the place \emph{answered} of the reservation system being an output place of each transition \emph{t1}, it will be marked after firing \emph{t1}. Thus, the number of tokens present at this place represents the number of positive answers received by the machines. When the
required number of resources (denoted here by $n$) is available ($n$ tokens are present in place \emph{answered}), the
\emph{launch} transition can fire producing $n$ tokens
in the place \emph{launching\_job}, corresponding to the action
of actually triggering the start of the application's execution. 

  \begin{center}
    \scalebox{.6}{\begin{tikzpicture}[node distance=1.3cm,>=stealth',auto]
  \node [place, tokens=1, label=begin] (begin0) {};
  \node [transition] (start0) [below = of begin0, label=right:start\_job, label=below right:{$[0,0]$}] {};
  \node [place, below =of start0, label=right:get\_nodes, label=below right:{$[0,0]$}] (getnodes0) {};

  \draw [->] (begin0) -- (start0);
  \draw [->] (start0) -- (getnodes0) node [midway] {$n$};

  \node [place, below=of getnodes0, label=above:answered] (answered0) {};

  \node [transition] (launch0) [below = of answered0, label=above:launch, label=right:{$[0,0]$}] {};
  \node [place, below =of launch0, label=left:launching\_job] (launching0) {};

  \node (points) [below = of launching0]{ \it \large . . . .};
  \node (res00) [left = of points]{ \it reserve};
  \node (res40) [right= of points]{ \it reserve};

  \draw [dashed,->] (getnodes0) -- (res00.north);
  \draw [dashed,->] (getnodes0) -- (res40.north);

  \draw [->] (answered0) -- (launch0) node [midway] {$n$};
  \draw [->] (launch0) -- (launching0) node [midway] {$n$};

  \draw [dashed,->] (res00) -- (answered0);
  \draw [dashed,->] (res40) -- (answered0);

  \node (can00) [below = 0mm of res00]{ \it cancel};
  \node (can40) [below = 0mm of res40]{ \it cancel};

  \draw [dashed,->] (answered0) -- (can00.east);
  \draw [dashed,->] (answered0) -- (can40.west);

  \node (start00) [below = of res00]{ \it start};
  \node (start40) [below = of res40]{ \it start};

  \draw [dashed,->] (launching0) -- (start00);
  \draw [dashed,->] (launching0) -- (start40);

    \draw[dashed] ($(res00.north west)+(-0.2,0.15)$) rectangle ($(start00.south east)+(0.3,-0.15)$);
\node [below of = start00] {\it machine};
    \draw[dashed] ($(res40.north west)+(-0.2,0.15)$) rectangle ($(start40.south east)+(0.3,-0.15)$);
\node [below of = start40] {\it machine};

\end{tikzpicture}	}
  
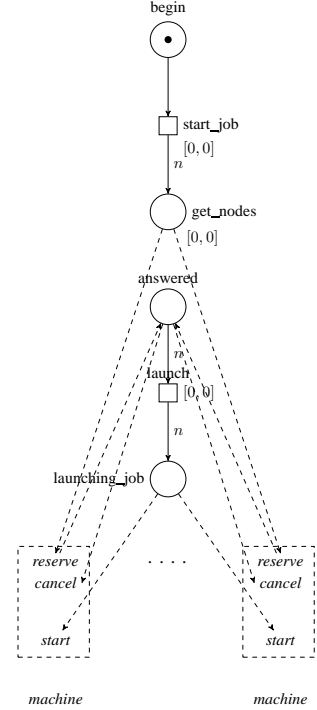
\captionof{figure}{\label{fig:qurdcli}Model of the reservation system}
  \end{center}

The place \emph{launching\_job} is an input place of each \emph{t2}
transition of the machine model. If a machine has been reserved for this
application, a token is present in its \emph{reserved} place. Therefore, the
\emph{t2} transition can be fired as soon as all the required ressources answered positively and each involved machine moves to state \emph{running} (represented by place \emph{running}). 

When the totality of the required ressources is not available, the system will be in a situation where $n_1$ tokens are in place \emph{answered} and $n_2$ tokens are in place  \emph{get\_nodes} such that $n_1 >0$, $n_2 \geq 0$ and $n_1 +n_2 =n$. In order to avoid holding these resources
indefinitely, the system must free them after some timeout (which is not explicitly represented in the model but could be by associating a time interval with transition \emph{cancel} as for Time Petri nets). When a machines stays at the state \emph{reserved} more than the timeout, the \emph{cancel} transition of
the machine is fired, consuming a token from the \emph{answered}
place. The machine is freed (it returns into state \emph{available}), it is removed from the list of resources that are considered by the client as having answered positively (by consuming the corresponding token from place \emph{answered}). Finally, this timeout mechanism can be used as well when the client disconnects from the system before the application is started (this is observed from the machine size by the fact that the application is not launched by the client after the timeout.

When the execution of a process is done, the machine exits from the state
\emph{running} and gets into state \emph{finished}. When a token is in the resources'
\emph{finished} place, transition \emph{t4} can be fired. The firing of  \emph{t4} produces two
tokens: one of them goes into the \emph{job\_finished} place, which
models the fact that this particular resource has signaled to the
client the fact that its process is done, and the other one goes into the
\emph{available} place, to put the machine back into the set of
machines that are available.


In section \ref{sec:algo:launcher} we mentioned two possible semantics
for the case when not enough resources are declared on the Zeroconf
bus at the moment when the client issues a request. In the \emph{fail
  semantics}, all the machines are released and the client does not
get any resource at all. In this case, all the reserved machines fire the
\emph{cancel} transition. In the \emph{wait semantics}, the client
waits until enough machines are available, until the timeout is
reached (in order to avoid deadlocks between concurrent
applications). Hence, both of these semantics are modeled.

\subsection{Modeling the Zeroconf publish/unpublish semantics}
\label{sec:model:zeroconf}

Additionally, for some reason a machine can withdraw itself from the
Zeroconf bus and publish itself back. For instance, if a service
suddenly takes all the memory of the server, there may not be enough
memory left to host new services. Or a machine can be turned off by an
administrator. An optional component of our model can describe this
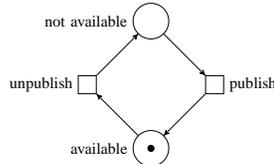
\begin{wrapfigure}{r}{4cm}
  \begin{center}
    \scalebox{.6}{\begin{tikzpicture}[node distance=1.3cm,>=stealth',auto]
  \node [place, tokens=1, label=left:available] (available0) {};
  \node [transition] (t100) [above left = of available0, label=left:unpublish] {};
  \node [place, above right = of t100, label=left:{not available}] (notavailable0) {};
  \node [transition] (t200) [above right= of available0, label=right:publish] {};
  \draw [->] (available0) -- (t100);
  \draw [->] (t100) -- (notavailable0);
  \draw [->] (notavailable0) -- (t200);
  \draw [->] (t200) -- (available0);

\end{tikzpicture}	}
  \captionof{figure}{\label{fig:zeroconf}Zeroconf semantics}
  \end{center}
\end{wrapfigure}
behavior. The model is represented by figure \ref{fig:zeroconf}. 

When a machine has declared itself on the Zeroconf bus, a token is
placed in its \emph{available} place. When it is unpublished from the
bus, the \emph{unpublish} transition is fired and a token is placed in
the \emph{not\_available} place. The only way it can be used is if
there is a token in the \emph{available} place. Hence, it needs to
fire the \emph{publish} transition to be used again, modeling the fact
that the machine has published itself back on the Zeroconf bus.

\subsection{Volatility}
\label{sec:model:volatile}

To be safe and robust, such a system must be able to handle
volatility. In this section, we present how both sides (client and
resources) are handled in a robust way in order to make sure that
failures or disappearances have no impact on the reservation system. 

\subsubsection{Client volatility}

Client volatility is due to the fact that a client can
leave the network unexpectedly at any moment. The model described in
section \ref{sec:model:qurdcli} already handles this fact. If a client
disappears when a machine is in state \emph{available}, it has no
incidence on this machine. If the client disappears after a
reservation has been issued (the machine is in state \emph{reserved}),
the timeout mechanism makes sure that the machines are freed and
become available for other applications. If the client disappears when the
machine is either in state \emph{running} or in state \emph{finished}, the application is
already running on the machines and the client can come back later to
collect the results with no incidence. 

\subsubsection{Resource volatility}

A resource can disappear while a process is running on this resource. Here
we assume that the system is using an \emph{eventually perfect}
failure detector, such as Heartbeat \cite{ACT}. We can model this
failure detector by a specific place in the system: the \emph{failure
  detector} place. While a machine is in state \emph{running}, its
disappearance is modeled by firing a \emph{dead} transition, which
puts a token in the place \emph{failure detector}. 

The failure detector is in charge with finding a new available machine
and restarting the failed process. If a machine has a token in its place
\emph{available}, it can fire transition \emph{restart} and
set this machine directly in state \emph{running}. It should be noted
that the \emph{restart} transitions are resource-specific, whereas the
failure detector is a global component. 

If a machine disappears while in state \emph{reserved}, the timeout
mechanism allows to fire transition \emph{cancel} (see Figure \ref{fig:fd}). 

\begin{center}
  \scalebox{.6}{\begin{tikzpicture}[node distance=1.3cm,>=stealth',auto]
  \node [place, tokens=1, label=available] (available0) {};
  \node [transition] (t10) [below = of available0, label=left:{t1 $[0,0]$}] {};
  \node [place, tokens=0, label=below right:reserved, below of=t10] (reserved0) {};
  \node [transition] (t20) [below of=reserved0, label=left:{t2 $[0,0]$}] {};
  \node [place, tokens=0, label=right:running, below of=t20] (running0) {};
  \node [transition] (t30) [below of=running0,label=left:{t3 $[T,T]$}] {};
  \node [place, tokens=0, label=right:finished, below of=t30] (finished0) {};
  \node [transition] (t40) [left=of finished0, label={left:t4 $[0,0]$}] {};
  \node [transition] (t50) [right=of reserved0, label=right:cancel, label=below right:{$[to,to]$}] {};

  \draw [->] (available0) -- (t10);
  \draw [->] (t10) -- (reserved0);
  \draw [->] (reserved0) -- (t20);
  \draw [->] (t20) -- (running0);
  \draw [->] (running0) -- (t30);
  \draw [->] (t30) -- (finished0);
  \draw [->] (finished0) -- (t40);
  \draw [->] (t40) -- (available0);
  \draw [->] (reserved0) -- (t50);
  \draw [->] (t50) -- (available0);
  
  \node (res0) [left = of available0]{ \it request};
  \draw [->, dashed] (res0) -- (t10);
  \node (done0) [below = of finished0]{ \it job\_finished};
  \draw [->, dashed] (finished0) -- (done0);
  \node (answ0) [right = of available0]{ \it answered};
  \draw [->, dashed] (t10) -- (answ0);
  \draw [->, dashed] (answ0) -- (t50);
  \node (launch0) [left = of reserved0]{ \it launch\_job};
  \draw [->, dashed] (launch0) -- (t20);

  \node [transition] (crash0) [right = of t30, label=right:{crash $[M,M]$}] {};

  \draw [->] (running0) -- (crash0);

  \node [place, below right = of crash0, xshift=3cm, label=below:failure detector] (fd0) {};
  \draw [->] (crash0) -- (fd0);
  \node [transition, right = of t50, xshift=1cm, label=right:{restart $[0,0]$}] (restart00) {};
  \draw [->] (available0) -- (restart00);
  \draw [->] (fd0) -- (restart00);
  \draw [->] (restart00) -- (running0);

  \node [place, right = of restart00, xshift=2cm, label=below:dead] (dead0) {};
  \draw [->] (crash0) -- (dead0);
  \node [transition, above = of dead0, label={rejuvenate $[T_r,T_r]$}] (rej0) {};
  \draw [->] (dead0) -- (rej0);
  \draw [->] (rej0) -- (available0);

\end{tikzpicture}	}	
  
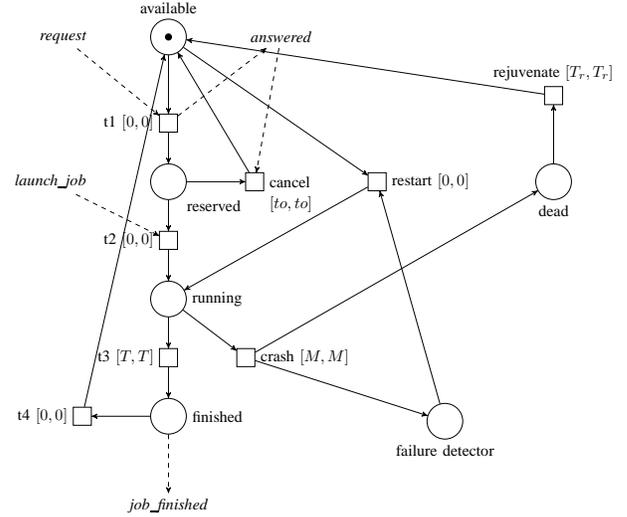
\captionof{figure}{\label{fig:fd}Model for handling resource volatility with a
    failure detector}
  \end{center}

\subsection{Concurrency between clients}
\label{sec:model:concurrent}

When several clients are issuing resource requests, there may be a
situation of concurrency between the clients. Each client browses the
Zeroconf bus and gets a list of available resources. A given resource
can be discovered by several clients in the same time. However, the fact
that a machine can answer positively only when it is in the state
\emph{available} (before switching to state \emph{reserved})
makes sure that it can answer positively to one client only. This is
modeled by the fact that there is only one \emph{available} place on
each resource, and this place contains only one token in the initial 
marking. 

A model where 2 clients issue concurrent resource allocation requests
on the same set of resources is represented on Figure~\ref{fig:resaconcur:client}. Each client has its own reservation system. On
the figure, we represented a client requesting $n$ resources and
another one requesting $m$ resources.

\begin{center}
  \resizebox{0.4\textwidth}{!}{\begin{tikzpicture}[node distance=1.3cm,>=stealth',auto]
  \node [place, tokens=1, label=begin0] (begin0) {};
  \node [transition] (start0) [below = of begin0, label=right:start\_job0, label=below right:{$[0,0]$}] {};
  \node [place, below =of start0, label=right:get\_nodes0] (getnodes0) {};

  \draw [->] (begin0) -- (start0);
  \draw [->] (start0) -- (getnodes0) node [midway] {$n$};

  \node [place, right= 3cm of begin0, tokens=1, label=begin1] (begin1) {};
  \node [transition] (start1) [below = of begin1, label=right:start\_job1, label=below right:{$[0,0]$}] {};
  \node [place, below =of start1, label=right:get\_nodes1] (getnodes1) {};

  \draw [->] (begin1) -- (start1);
  \draw [->] (start1) -- (getnodes1) node [midway] {$m$};

  \node [place, below=of getnodes0, label=above:answered0] (answered0) {};

  \node [transition] (launch0) [below = of answered0, label=left:launch0, label=below left:{$[0,0]$}] {};
  \node [place, below =of launch0, label=left:launching\_job0] (launching0) {};

  \node [place, below=of getnodes1, label=above:answered1] (answered1) {};

  \node [transition] (launch1) [below = of answered1, label=left:launch1, label=below left:{$[0,0]$}] {};
  \node [place, below =of launch1, label=left:launching\_job1] (launching1) {};

  \node (res00) [below left = of launching0]{ \it reserve0};
  \node (res01) [below = 1mm of res00]{ \it reserve1};

  \node (res40) [below right = of launching1]{ \it reserve0};
  \node (res41) [below = 1 mm of res40]{ \it reserve1};

  \draw [dashed,->] (getnodes0) -- (res00);
  \draw [dashed,->] (getnodes0) -- (res40);

  \draw [dashed,->] (getnodes1) -- (res01);
  \draw [dashed,->] (getnodes1) -- (res41);

  \draw [->] (answered0) -- (launch0) node [midway] {$n$};
  \draw [->] (launch0) -- (launching0) node [midway] {$n$};
  \draw [->] (answered1) -- (launch1) node [midway] {$m$};
  \draw [->] (launch1) -- (launching1) node [midway] {$m$};

  \draw [dashed,->] (res00) -- (answered0);
  \draw [dashed,->] (res40) -- (answered0);
  \draw [dashed,->] (res01) -- (answered1);
  \draw [dashed,->] (res41) -- (answered1);

  \node (start00) [below = of res01]{ \it start0};
  \node (start40) [below = of res41]{ \it start0};
  \node (start01) [below = 1mm of start00]{ \it start1};
  \node (start41) [below = 1mm of start40]{ \it start1};

  \draw [dashed,->] (launching0) -- (start00);
  \draw [dashed,->] (launching0) -- (start40);
  \draw [dashed,->] (launching1) -- (start01);
  \draw [dashed,->] (launching1) -- (start41);

  \node (finished00) [below = of start01]{ \it finished0};
  \node (finished40) [below = of start41]{ \it finished0};
  \node (finished01) [below = 1mm of finished00]{ \it finished1};
  \node (finished41) [below = 1mm of finished40]{ \it finished1};

  \node (points) [right= 2cm of start01]{ \it \large . . . .};

    \draw[dashed] ($(res00.north west)+(-0.2,0.15)$) rectangle ($(finished01.south east)+(0.3,-0.15)$);
\node (machine0) [left= of finished01] {\it machine};
    \draw[dashed] ($(res40.north west)+(-0.2,0.15)$) rectangle ($(finished41.south east)+(0.3,-0.15)$);
\node [left = of finished41] {\it machine};

  \node [place, below=of finished01, label=left:job\_finished0] (jobfinished0) {};
  \node [place, below=of finished41, label=left:job\_finished1] (jobfinished1) {};

  \draw [dashed,->] (finished00) -- (jobfinished0);
  \draw [dashed,->] (finished01) -- (jobfinished1);
  \draw [dashed,->] (finished40) -- (jobfinished0);
  \draw [dashed,->] (finished41) -- (jobfinished1);

\end{tikzpicture}	}
  
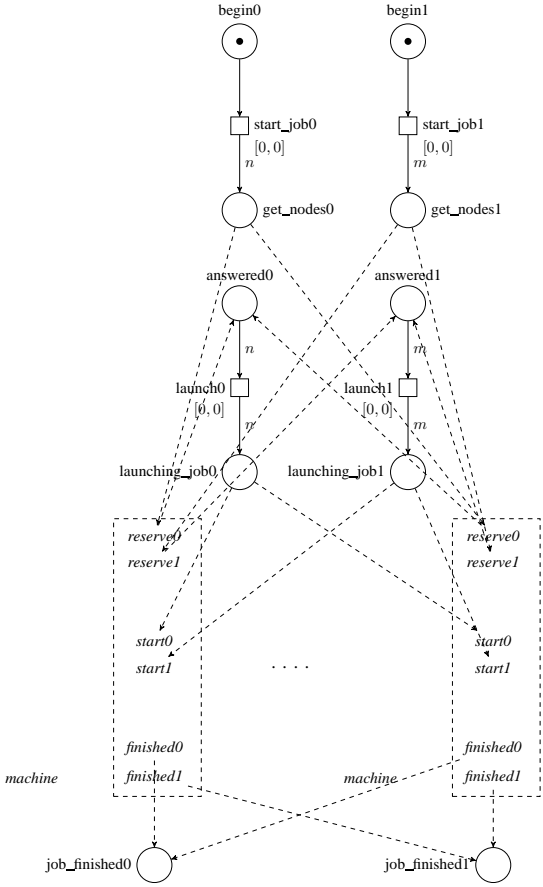
\captionof{figure}{Reservation system of two clients on a given set
    of resources\label{fig:resaconcur:client}}
\end{center}

The local state of each resource depends on the
application it is working for. For instance, the state \emph{reserved}
is specific to ``application X''; hence, the state is actually
\emph{reserved for application X}. Therefore, the execution path of
the model from the reservation to the end of the execution of the
local process must be specific for each application. However, as
explained earlier in this section, since the initial marking features
only one token in place \emph{available} of the machine, only one
of these paths can be active at a time. The model for a machine that may answer to two applications is
represented in figure \ref{fig:resaconcur:machine}. The left path
correspond to an application called 0; the right path corresponds to
an application called 1. When the resource is in state \emph{available}
 and application 0 issues a request, the \emph{t10} transition is
fireable resulting for the resource to turn into state \emph{reserved for 0} 
and there  is no token left in place \emph{available}. As a
consequence, if application 1 issues a request, the \emph{t11}
transition cannot be fired and as long as the resource is not back
into \emph{available} mode, it will not be able to be used by
application 1.

The \emph{cancel} transition is very important here to release some
resources in case of a deadlock caused by a conflict between
applications in the wait semantics (see section
\ref{sec:algo:launcher}, Algorithm \ref{algo:reservationWait}). For
instance, if we have three resources and two applications asking for three and two machines respectively, then the following deadlock state can be reached: The first application gets two machines and the second one gets one machine. Thus, all the available machines are reserved but no application is able to start. Therefore, after a
certain time, if no additional resources appear on the Zeroconf bus, the machines reserved for at least one application will be freed and become available for the other one.

\begin{center}
  \resizebox{0.4\textwidth}{!}{\begin{tikzpicture}[node distance=1.3cm,>=stealth',auto]
  \node [place, tokens=1, label=available] (available0) {};

  \node [transition] (t10) [below left = 3cm of available0, label=right:t10] {};
  \node [place, tokens=0, label=right:reserved0, below of=t10] (reserved0) {};
  \node [transition] (t20) [below of=reserved0, label=right:t20] {};
  \node [place, tokens=0, label=left:running0, below of=t20] (running0) {};
  \node [transition] (t30) [below of=running0,label=right:t30] {};
  \node [place, tokens=0, label=left:finished0, below of=t30] (finished0) {};
  \node [transition] (t40) [right=of finished0, label=below:t40] {};
  \node [transition] (t50) [left=of reserved0, label=below:cancel0] {};
 
  \draw [->] (available0) -- (t10);
  \draw [->] (t10) -- (reserved0);
  \draw [->] (reserved0) -- (t20);
  \draw [->] (t20) -- (running0);
  \draw [->] (running0) -- (t30);
  \draw [->] (t30) -- (finished0);
  \draw [->] (finished0) -- (t40);
  \draw [->] (t40) -- (available0);
  \draw [->] (reserved0) -- (t50);
  \draw [->] (t50) -- (available0);
  
  \node (res0) [left = of available0]{ \it request0};
  \draw [->, dashed] (res0) -- (t10);
  \node (done0) [below = of finished0]{ \it job\_finished0};
  \draw [->, dashed] (finished0) -- (done0);
  \node (answ0) [left = of res0]{ \it answered0};
  \draw [->, dashed] (t10) -- (answ0);
  \draw [->, dashed] (answ0) -- (t50);
  \node (launch0) [below = of answ0]{ \it launch\_job0};
  \draw [->, dashed] (launch0) -- (t20);

  \node [transition] (t11) [below right= 3cm of available0, label=left:t11] {};
  \node [place, tokens=0, label=left:reserved1, below of=t11] (reserved1) {};
  \node [transition] (t21) [below of=reserved1, label=left:t21] {};
  \node [place, tokens=0, label=right:running1, below of=t21] (running1) {};
  \node [transition] (t31) [below of=running1,label=left:t31] {};
  \node [place, tokens=0, label=right:finished1, below of=t31] (finished1) {};
  \node [transition] (t41) [left=of finished1, label=below:t41] {};
  \node [transition] (t51) [right=of reserved1, label=below:cancel1] {};

  \draw [->] (available0) -- (t11);
  \draw [->] (t11) -- (reserved1);
  \draw [->] (reserved1) -- (t21);
  \draw [->] (t21) -- (running1);
  \draw [->] (running1) -- (t31);
  \draw [->] (t31) -- (finished1);
  \draw [->] (finished1) -- (t41);
  \draw [->] (t41) -- (available0);
  \draw [->] (reserved1) -- (t51);
  \draw [->] (t51) -- (available0);
  
  \node (res1) [right = of available0]{ \it request1};
  \draw [->, dashed] (res1) -- (t11);
  \node (done1) [below = of finished1]{ \it job\_finished1};
  \draw [->, dashed] (finished1) -- (done1);
  \node (answ1) [right= of res1]{ \it answered1};
  \draw [->, dashed] (t11) -- (answ1);
  \draw [->, dashed] (answ1) -- (t51);
  \node (launch1) [below = of answ1]{ \it launch\_job1};
  \draw [->, dashed] (launch1) -- (t21);

\end{tikzpicture}	}
  
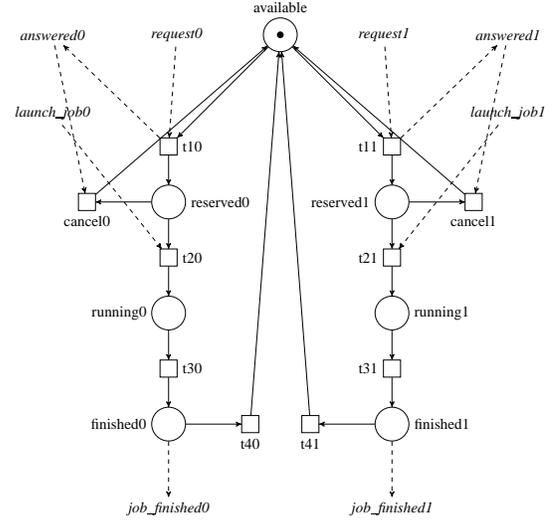
\captionof{figure}{Two concurrent requests on a single resource\label{fig:resaconcur:machine}}
\end{center}

\section{Complete models}
\label{sec:complet}

\section{Conclusion}

In this report, we have described a model using Petri Nets for a
decentralized resource reservation system. This model can now be used
to verify properties regarding the behavior of this system. For
instance, one can verify under which conditions all the submitted jobs
can complete. 

\footnotesize
\bibliographystyle{plain}
\bibliography{qurd}

\end{document}